\ifdefined\anonymous
\documentclass[sigconf,nonacm,anonymous]{acmart}
\else
\documentclass[sigconf,nonacm]{acmart}
\fi

\setcopyright{acmcopyright}
\copyrightyear{2024}
\acmYear{2024}
\acmDOI{XXXXXXX.XXXXXXX}

\acmConference[DAC '24]{DAC '24}{June 23--27, 2024}{San Francisco, CA}
\acmBooktitle{Design Automation Conference, June 23-27, 2024, San Francisco, CA}
\acmPrice{15.00}
\acmISBN{978-1-4503-XXXX-X/18/06}

\renewcommand{\baselinestretch}{0.961}

\usepackage{xcolor}
\usepackage{amsfonts} 
\usepackage{textcomp}
\usepackage[free-standing-units,per-mode=symbol,mode=text,reset-text-series=false,reset-text-family=false,text-series-to-math=true,text-family-to-math=true]{siunitx}
\usepackage{tikz} 
\usepackage{glossaries}
\usepackage{import} 
\usepackage{microtype}
\usepackage{url}
\usepackage[inline]{enumitem}
\usepackage{multirow}
\usepackage{booktabs}
\usepackage[capitalise,noabbrev,nameinlink]{cleveref}
\usepackage[position=b]{subcaption}
\usepackage[frozencache,cachedir=minted]{minted}
\usepackage[export]{adjustbox}
\usepackage{hyphenat}

\acmSubmissionID{1403}

\glsdisablehyper

\newacronym{hpc}{HPC}{high-performance computing}
\newacronym{isa}{ISA}{instruction set architecture}
\newacronym{fpu}{FPU}{floating-point unit}
\newacronym{sssr}{SSSR}{sparse stream semantic register}
\newacronym{frep}{FREP}{floating-point repetition}
\newacronym{rtl}{RTL}{register transfer level}
\newacronym{mac}{MAC}{multiply-accumulate}
\newacronym{sr}{SR}{stream register}
\newacronym{tcdm}{TCDM}{tightly coupled data memory}
\newacronym{dma}{DMA}{direct memory access}
\newacronym{flop}{FLOP}{floating-point operation}
\newacronym{ipc}{IPC}{instructions per cycle}
\newacronym[firstplural=systems-on-chip (SoCs)]{soc}{SoC}{system-on-chip}
\newacronym{wse}{WSE}{wafer-scale engine}
\newacronym{fpga}{FPGA}{field-programmable gate array}
\newacronym{llc}{LLC}{last-level cache}
\newacronym{sm}{SM}{shared memory}
\newacronym{dag}{DAG}{directed acyclic graph}
\newacronym{dlt}{DLT}{data layout transform}
\newacronym{tlb}{TLB}{translation lookaside buffer}
\newacronym{cmtr}{CMTR}{compute-to-memory time ratio}

\newcommand{\x}{$\times$}

\ifdefined\valcheck
\newcommand{\impval}[1]{\noindent{\color{olive}{#1}}}
\else
\newcommand{\impval}[1]{#1}
\fi

\DeclareSIUnit{\nothing}{\relax}
\DeclareSIUnit{\gateequi}{GE}

\DeclareSIUnit{\bit}{b}
\DeclareSIUnit{\flop}{FLOP}

\usepackage{listings}
\AtBeginDocument{\DeclareCaptionSubType{lstlisting}}
\crefname{sublstlisting}{listing}{listings}
\Crefname{sublstlisting}{Listing}{Listings}

\newcommand{\topar}[1]{\textit{#1}:}

\begin{document}


\title{SARIS: Accelerating Stencil Computations on Energy-Efficient RISC-V Compute Clusters with Indirect Stream Registers}

\author{Paul Scheffler}
\email{paulsc@iis.ee.ethz.ch}
\affiliation{%
  \institution{Integrated Systems Lab., ETH Zurich}
  \city{Zurich}
  \country{Switzerland}
}

\author{Luca Colagrande}
\email{colluca@iis.ee.ethz.ch}
\affiliation{%
  \institution{Integrated Systems Lab., ETH Zurich}
  \city{Zurich}
  \country{Switzerland}
}

\author{Luca Benini}
\email{lbenini@iis.ee.ethz.ch}
\affiliation{%
  \institution{Integrated Systems Lab., ETH Zurich}
  \city{Zurich}
  \country{Switzerland}
}
\affiliation{%
  \institution{University of Bologna}
  \city{Bologna}
  \country{Italy}
}



\begin{abstract}
Stencil codes are performance-critical in many compute-intensive applications, but suffer from significant address calculation and irregular memory access overheads. This work presents SARIS, a general and highly flexible methodology for stencil acceleration using register-mapped indirect streams. We demonstrate SARIS for various stencil codes on an eight-core RISC-V compute cluster with indirect stream registers, achieving significant speedups of \impval{2.72x}, near-ideal FPU utilizations of \impval{81\%}, and energy efficiency improvements of \impval{1.58x} over an RV32G baseline on average. Scaling out to a  256-core manycore system, we estimate an average FPU utilization of \impval{64\%}, an average speedup of \impval{2.14x}, and up to \impval{15\%} higher fractions of peak compute than a leading GPU code generator.  
\end{abstract}



\ifdefined\anonymous
\begin{teaserfigure}
\vspace{-0.1em}
\end{teaserfigure}
\fi

\maketitle


\section{Introduction}
\label{sec:introduction}

Stencil codes are performance-critical in numerous data-parallel, compute-intensive applications in domains like physical simulation, signal processing, and machine learning.
They iterate over points in multi-dimensional data grids and update them based on their neighbors in a fixed pattern called a \emph{stencil}.

Stencil codes share a common structure, but are highly diverse in their computational profiles, ranging from simple image filters~\cite{Pouchet2012Polybench} to complex seismic propagation operators~\cite{Jacquelin2022ScalableDH}.
The number and dimensions of I/O arrays, operations per point, and stencil shape all vary significantly between codes~\cite{Rawat2019OnOptimizingCS}, making their acceleration on modern parallel processors challenging. 
Existing software proposals accelerating stencils present both generic solutions, such as code generators~\cite{Rawat2019OnOptimizingCS, You2021DRStencilED, Zhang2023RevisitingTB, Matsumura2020AN5DAS} or platform optimizations~\cite{Zhang2020DataLT, Yount2015VectorFI, Zhao2018DeliveringPS}, and tuned implementations of specific codes~\cite{Jacquelin2022ScalableDH, Rocki2020FastSC}. 
In both cases, the target platform's architectural features are heavily leveraged.

One major source of remaining inefficiencies are \emph{memory accesses}. 
For instance, small stencils tend to result in low operational intensities and \emph{memory-bound} execution~\cite{Singh2020NEROAN, Rawat2019OnOptimizingCS, Zhao2018DeliveringPS}. 
Codes with irregular stencil shapes or many I/O arrays suffer from \emph{irregular access patterns} inefficiently handled by tiered memory hierarchies~\cite{Denzler2021CasperAS, Singh2020NEROAN, Yount2015VectorFI}. 
Finally, both small and irregular stencils introduce significant address calculation and load-store overheads, degrading performance particularly in energy-efficient single-issue, in-order processors.

\begin{figure}[t]
  \centering
  \includegraphics[height=6.2em]{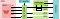}
  \caption{Integration of three floating-point \glspl{sr}. 
  When configured, the reg decode block maps accesses to registers associated with streams to SRs, which act as FIFO interfaces to memory. Addresses are produced by hardware generators.}
  \label{fig:intro_sr}
\end{figure}

Recently, a class of hardware extensions tackling these memory inefficiencies has emerged: 
\emph{stream registers} (\glsunset{sr}\glspl{sr})~\cite{Schuiki2021StreamSR, Wang2019StreambasedMA, Scheffler2023SparseSS, Domingos2021UnlimitedVE} map streams of memory accesses directly to reads or writes of architectural registers, with addresses generated by a dedicated hardware unit.
A simple integration of three floating-point \glspl{sr} is shown in \Cref{fig:intro_sr}.
\glspl{sr} enable continuous streaming of useful data, %
maximizing bandwidth utilization and thus performance on memory-bound workloads. 
They also decouple memory access from computation and free the processor from handling loads, stores, and address generation, enabling near-ideal \gls{fpu} utilizations in data-intensive workloads even on single-issue, in-order cores~\cite{Schuiki2021StreamSR, Scheffler2023SparseSS}.
Many \glspl{sr} also support \emph{indirect} streams~\cite{Wang2019StreambasedMA, Scheffler2023SparseSS, Domingos2021UnlimitedVE} using a base address and index arrays to scatter or gather data, which can accelerate \emph{arbitrarily irregular} access patterns.

In this paper, we present \emph{SARIS}, a generic methodology for \textbf{S}tencil \textbf{A}cceleration using \textbf{R}egister-mapped \textbf{I}ndirect \textbf{S}treams.
SARIS encodes the offsets of grid elements accessed in the loop body of stencil codes in index arrays; it then reuses these indices on each point update, using the point's coordinates as an indirection base.
SARIS is highly flexible, applicable to \emph{any} stencil shape, and amenable to parallelization. 
It leverages concurrent streams through multiple indirect \glspl{sr} and additional affine \glspl{sr} where beneficial. 
It is orthogonal to existing code optimizations including arithmetic reassociation, loop unrolling, and the use of hardware loops.

{\renewcommand\fcolorbox[4][]{\textcolor{cyan}{\strut#4}}
\begin{figure*}[ht!]
\centering%
\begin{minipage}{0.31\textwidth}%
\centering%
\subfloat[Visualization of stencil shape, constant parameters, and sweep over 3D grid.]{%
\hspace{1.8em}
\includegraphics[width=0.72\linewidth,valign=t]{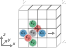}%
\label{fig:impl_saris_ex_stc}%
\hspace{1.8em}
}
\vspace{1.9em}
\subfloat[SARIS point loop schedule showing register streams and compute operations.]{%
\includegraphics[width=\linewidth,valign=b]{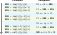}%
\label{fig:impl_saris_ex_feed}%
}%
\caption{Visualization and schedule for the symmetric 7-point star stencil code.}
\label{fig:impl_saris_ex}
\end{minipage}
\hspace{1em}
\begin{minipage}{0.66\textwidth}
\centering
\setcaptiontype{lstlisting}
\begin{minipage}[t]{0.4965\linewidth}
\begin{minipage}[t]{\linewidth}
\begin{minted}[fontsize=\fontsize{6.6}{7.0}\selectfont]{python}
for z in 1 to N-1:
  for y in 1 to N-1:
    for x in 1 to N-1:
      out[z][y][x] =
          c0 *    inp[z][y][x]
        + cx * (  inp[z][y][x-1]
                + inp[z][y][x+1])
        + cy * (  inp[z][y-1][x]
                + inp[z][y+1][x])
        + cz * (  inp[z-1][y][x]
                + inp[z+1][y][x]);
\end{minted}
\vspace{-0.3em}
\subcaption{Baseline time iteration pseudocode.}
\label{lst:impl_saris_ex_basehl}%
\vspace{0.75em}
\end{minipage}\\
\begin{minipage}[b]{\linewidth}
\begin{minted}[fontsize=\fontsize{6.6}{6.8}\selectfont, baselinestretch=0.91]{gas}
x:  fld     ft0, 0(t0)      # inp[z][y][x]
    fmul.d  ft0, %[c0], ft0
    fld     ft1, -8(t0)     # inp[z][y][x-1]
    fld     ft2,  8(t0)     # inp[z][y][x+1]
    fadd.d  ft1, ft1, ft2
    fmadd.d ft0, %[cx], ft1, ft0
    fld     ft1, -YOFFS(t0) # inp[z][y-1][x]
    fld     ft2,  YOFFS(t0) # inp[z][y+1][x]
    fadd.d  ft1, ft1, ft2
    fmadd.d ft0, %[cy], ft1, ft0
    fld     ft1, 0(t1)      # inp[z-1][y][x]
    fld     ft2, 0(t2)      # inp[z+1][y][x]
    fadd.d  ft1, ft1, ft2
    fmadd.d ft0, %[cz], ft1, ft0
    fsd     ft0, 0(t3)
    addi    t0, 8
    addi    t1, 8
    addi    t2, 8
    addi    t3, 8
    bne     t0, a0, x
\end{minted}
\vspace{-0.3em}
\subcaption{Baseline point loop RISC-V assembly.}
\label{lst:impl_saris_ex_baseasm}%
\end{minipage}%
\end{minipage}
\begin{minipage}[t]{0.4965\linewidth}
\centering
\begin{minipage}[t]{\linewidth}
\begin{minted}[fontsize=\fontsize{6.6}{7.0}\selectfont]{python}
# Configure input offsets as indices (SR 0,1).
sr_set_idcs(SR0, {&inp[1][1][1], &inp[1][1][2],
                  &inp[1][2][1], &inp[2][1][1]});
sr_set_idcs(SR1, {&inp[1][1][0], &inp[1][0][1],
                  &inp[0][1][1] });
#Affine write stream stores outputs (SR 2).
sr_affine_write_3d(SR2, &out[1][1][1], ...);

for z in 0 to N-2:
  for y in 0 to N-2:
    for x in 0 to N-2;
      # Launch indirect input reads with
      # point offset as array base (SR 0,1)
      sr_indir_read(SR0|SR1, OFFS(x,y,z), ...);
      # Computation in order of point loop sched.
      acc =  c0 * sr_read(SR0);
      acc += cx * (sr_read(SR0) + sr_read(SR1));
      acc += cy * (sr_read(SR0) + sr_read(SR1));
      acc += cz * (sr_read(SR0) + sr_read(SR1));
      sr_write(SR2, acc);
\end{minted}
\vspace{-0.3em}
\subcaption{SARIS time iteration pseudocode.}
\label{lst:impl_saris_ex_sarishl}%
\vspace{0.7em}
\end{minipage}\\
\begin{minipage}[b]{\linewidth}
\begin{minted}[fontsize=\fontsize{6.6}{6.8}\selectfont]{gas}
x:  SRIR    SR0|SR1, t0     # SSSRs: 3 insts
    fmul.d  ft0, %[c0], SR0
    fadd.d  ft1, SR0, SR1
    fmadd.d ft0, %[cx], ft1, ft0
    fadd.d  ft1, SR0, SR1
    fmadd.d ft0, %[cy], ft1, ft0
    fadd.d  ft1, SR0, SR1
    fmadd.d SR2, %[cz], ft1, ft0
    addi    t0, 8
    bne     t0, a0, x
\end{minted}
\vspace{-0.3em}
\subcaption{SARIS point loop RISC-V assembly.}
\label{lst:impl_saris_ex_sarisasm}%
\end{minipage}\\
\end{minipage}
\captionof{listing}{Time iteration pseudocode and point loop RISC-V assembly for the symmetric 7-point star stencil code with and without SARIS.}
\label{lst:impl_saris_ex}
\end{minipage}
\end{figure*}

We use SARIS to optimize parallel stencil codes for the open-source, energy-efficient RISC-V Snitch compute cluster~\cite{Zaruba2020SnitchAT}, which features eight RV32G cores extended with \glspl{sssr} \cite{Scheffler2023SparseSS} and the \glsunset{frep}\gls{frep} hardware loop. 
We implement both platform-optimized RISC-V baseline codes and variants using SARIS to leverage the \gls{sssr} and \gls{frep} extensions,
which we provide free and open-source\footnote{\url{https://github.com/pulp-platform/snitch_cluster/tree/main/sw/saris}}.
Comparing our optimized baseline and SARIS-accelerated codes in cycle-accurate simulation, we find significant speedups of \impval{2.72\x}, near-ideal \gls{fpu} utilizations of \impval{\SI{81}{\percent}}, and energy efficiency improvements of \impval{1.58\x} on average. 
We also estimate the performance benefits of SARIS on a 256-core manycore. 
We achieve a good mean \gls{fpu} utilization of \impval{\SI{64}{\percent}}, a mean speedup of \impval{2.14\x}. We obtain up to \impval{\SI{15}{\percent}} higher fractions of peak compute than a leading GPU code generator even when considering the bandwidth and latency nonidealities of a complex memory system.
Our contributions are:

\begin{itemize}
\item We present SARIS, a generic and highly flexible approach to accelerating stencil codes with register-mapped indirect streams orthogonal to existing code optimizations. 
\item We implement optimized RISC-V baseline and SARIS\hyp{}accelerated parallel implementations of various stencil codes on the Snitch cluster with \glspl{sssr} and \gls{frep} extensions.
\item We evaluate our SARIS-accelerated codes in cycle-accurate simulation, achieving significant speedups of \impval{2.72\x}~over baseline codes, near-ideal \gls{fpu} utilizations of \impval{\SI{81}{\percent}}, and energy efficiency improvements of \impval{1.58\x}~ on average.
\item We estimate the performance benefits of SARIS on a 256-core manycore with a bandwidth-limiting HBM2E memory stack, finding an average \gls{fpu} utilization of \impval{\SI{64}{\percent}}, an average speedup of \impval{2.14\x}, and up to \impval{\SI{15}{\percent}} higher fractions of peak compute than a leading GPU code generator.
\end{itemize}


\section{Implementation}
\label{sec:impl}

We first describe our SARIS method (\Cref{subsec:impl_saris}) and how it can be combined with existing code optimizations (\Cref{subsec:impl_opt}). We then present our implementation of optimized baseline and SARIS-accelerated stencil codes on the Snitch cluster (\Cref{subsec:impl_snitch}).

\subsection{SARIS Method}
\label{subsec:impl_saris}

We describe SARIS using the example of a symmetric 7-point star stencil iterating over a single data array as shown in \Cref{fig:impl_saris_ex_stc}. 
We assume double-precision floating-point data, alternating buffers, and a grid halo initialized to sensible boundary conditions.

\Cref{lst:impl_saris_ex_basehl} shows the baseline pseudocode for one time iteration, where \texttt{inp} represents the current and \texttt{out} the next iteration buffer. 
Compiling the innermost point loop for the RV32G architecture%
\footnote{We assume here our grid tile is small enough for $y$ neighbors to be addressed using \SI{12}{\bit} immediate offsets, but too big to do this for $z$ neighbors, i.e. $16\leq N\leq 128$.} 
without further optimizations yields the assembly code in \Cref{lst:impl_saris_ex_baseasm}. 
Out of \impval{20} loop instructions, only \impval{7} (\impval{\SI{35}{\percent}}) do useful compute, while \impval{12} (\impval{\SI{60}{\percent}}) are dedicated to memory accesses and address calculation. 
As a result, \gls{fpu} utilization is limited to \impval{\SI{35}{\percent}} of its peak on energy-efficient single-issue in-order cores\footnote{While multi-issue out-of-order cores could further increase \gls{fpu} utilization through-instruction level parallelism, the associated energy overheads would be significant.}, even when ignoring stalls due to data dependencies and irregular memory access patterns. 
While unrolling and register-tiling data along the $x$ axis could eliminate up to \impval{two} loads per output point, this would still yield at most \impval{\SI{39}{\percent}} \gls{fpu} utilization.
As we will show in \Cref{subsec:eval_perf}, this limited compute efficiency manifests across all considered stencil codes.

The SARIS method accelerates stencil computation by minimizing the above memory access overheads through the use of \glspl{sr}. It involves the following steps on the point loop: 

\begin{enumerate}
    \item Map all \emph{grid data loads} to indirect stream reads.
    \item \emph{Partition} these reads among available indirect \glspl{sr}, maximizing their concurrent use and balancing their utilization.
    \item Map \emph{grid data stores} or loads of \emph{constant stencil coefficients} that cannot be kept in the register file to remaining \glspl{sr}.
    \item Determine a \emph{point loop schedule} specifying in which order the computations in the point loop access streams; this will determine the index arrays for each indirect \gls{sr}.
\end{enumerate}

The method may be iterated on for best performance, especially when combined with orthogonal optimizations like those described in \Cref{subsec:impl_opt}.
We demonstrate its steps on our 7-point-stencil code example;
as on our evaluation platform, we assume the two indirect \glspl{sr} (\texttt{SR0} and \texttt{SR1}) and one affine \gls{sr} (\texttt{SR2}) to be available:
\begin{enumerate}
    \item We map all seven grid data loads to indirect stream reads.
    \item For each axis, we map the two opposing grid point loads to \texttt{SR0} and \texttt{SR1}, respectively, so they can concurrently be read by an addition operation. The remaining center point load is mapped to \texttt{SR0}, which already results in minimal utilization imbalance between \texttt{SR0} and \texttt{SR1}.
    \item Since all four stencil coefficients ($c_0$, $c_x$, $c_y$, $c_z$) can be kept in the register file, we map our single grid data store to \texttt{SR2} using an affine 3D pattern along the grid.
    \item We use the same computation order as our baseline. The resulting point loop schedule shown in \Cref{fig:impl_saris_ex_feed} lists each compute operation and its stream accesses in order.
\end{enumerate}

The SARIS-accelerated pseudocode using our new point loop schedule is shown in \Cref{lst:impl_saris_ex_sarishl}. 
We first configure the static index arrays for \texttt{SR0} and \texttt{SR1}; for simplicity, we keep all indices positive by defining offsets around the iteration origin $(1,1,1)$ and changing the axis iteration ranges to $0\;..\;N-2$. 
We then configure \texttt{SR2} with the 3D iteration needed to write our output data.
Inside the point loop, we launch \texttt{SR0} and \texttt{SR1} with the grid point's address offset as the base; the indices remain the same for each grid point iteration. 
Finally, we perform our computation, reading and writing \glspl{sr} in the order defined by our point loop schedule.

\Cref{lst:impl_saris_ex_sarisasm} shows the resulting RISC-V assembly for the SARIS-accelerated point loop without further optimizations. %
Except for launch of \texttt{SR0} and \texttt{SR1} (\impval{3} instructions on Snitch with \glspl{sssr}), incrementing the grid point, and the loop branch, \emph{all} loop instructions now perform useful compute, almost doubling the ratio of useful compute instructions from \impval{\SI{35}{\percent}} to \impval{\SI{58}{\percent}}. %
Unlike for the baseline, however, the non-compute overhead in the point loop is \emph{static} and can significantly be reduced through complementary optimizations, enabling our near-ideal mean \gls{fpu} utilizations of \impval{\SI{81}{\percent}} in \cref{subsec:eval_perf}.

Going beyond our example, SARIS can accelerate \emph{any} stencil code as it makes very few assumptions. %
It supports any sequence of computations on grids of any dimensionality and size. %
The indirect streams enable arbitrarily shaped stencils, and since the indices include array bases, any number of I/O arrays may be streamed. %
Indices can even dynamically be rewritten if necessary.

In principle, SARIS could automatically be performed by the compiler whenever a fixed sequence of address offsets is repeatedly accessed with changing bases. This generalized formulation could also enable its application beyond stencil codes.
In this work, we focus on the manual application of the SARIS methodology to accelerate stencil codes as it is routinely done when optimizing low-level library code;
compiler inference is left as future work. 

\subsection{Complementary Optimizations}
\label{subsec:impl_opt}

SARIS is highly flexible; to further increase \gls{fpu} utilization, it can be combined with existing code optimizations: %

\topar{Unrolling} by unrolling the point loop and processing blocks of points at once, the impact of SARIS' static overheads can significantly be reduced. To this end, \gls{sr} streams can be extended to access grid points for multiple iterations. This also enables multi-dimensional unrolls, which can reduce parallelization imbalance.

\topar{Reordering and reassociation} by reordering indirect stream indices, point loop computations can freely be reordered and reassociated. This can improve performance by avoiding memory and dependency stalls, especially when combined with unrolling.

\topar{Hardware loops} SARIS is orthogonal to hardware loops, which reduce point iteration overheads. If they provide a separate instruction buffer, they also reduce instruction cache pressure and misses.

\topar{Parallelization} SARIS is compatible with parallelization at the grid point level. As every point iteration launches independent indirect streams, (unrolled blocks of) points can freely be distributed among multiple cores, for example in an interleaved fashion. Additionally, grids can be subdivided into tiles as usual.

\subsection{Stencils on a Snitch Cluster}
\label{subsec:impl_snitch}

\begin{table}[t!]
\centering
\small
\def\arraystretch{1.0}%
\begin{tabular}{p{9.2em}crrrr}
\toprule
  Code%
& Dims.%
& Rad.%
& \#Loads%
& \#Coeffs.%
& \#FLOPs%
\\ \midrule
\texttt{jacobi\_2d}~\cite{Pouchet2012Polybench}     & 2D & 1  & 5   & 1   & 5   \\
\texttt{j2d5pt}~\cite{Matsumura2020AN5DAS}          & 2D & 1  & 5   & 6   & 10  \\
\texttt{box2d1r}~\cite{Matsumura2020AN5DAS}         & 2D & 1  & 9   & 9   & 17  \\
\texttt{j2d9pt}~\cite{Matsumura2020AN5DAS}          & 2D & 2  & 9   & 10  & 18  \\
\texttt{j2d9pt\_gol}~\cite{Matsumura2020AN5DAS}     & 2D & 1  & 9   & 10  & 18  \\
\texttt{star2d3r}~\cite{Matsumura2020AN5DAS}        & 2D & 3  & 13  & 13  & 25  \\
\texttt{star3d2r}~\cite{Matsumura2020AN5DAS}        & 3D & 2  & 13  & 13  & 25  \\
\texttt{ac\_iso\_cd}~\cite{Jacquelin2022ScalableDH} & 3D & 4  & 26  & 13  & 38  \\
\texttt{box3d1r}~\cite{Matsumura2020AN5DAS}         & 3D & 1  & 27  & 27  & 53  \\
\texttt{j3d27pt}~\cite{Matsumura2020AN5DAS}         & 3D & 1  & 27  & 28  & 54  \\
\bottomrule
\end{tabular}
\vspace{1em}
\caption{Implemented stencil codes sorted by FLOPs per grid point; grid loads and coefficients are also per grid point.}
\vspace{-1em}
\label{tbl:impl_saris_codes}
\end{table}

We evaluate SARIS on the open-source, energy-efficient RISC-V Snitch cluster~\cite{Zaruba2020SnitchAT}, which provides eight single-issue, in-order RV32G cores. 
Each core consists of a minimal integer processor offloading instructions to a double-precision \gls{fpu}, whose utilization can be maximized through two cooperating \gls{isa} extensions.
The \emph{\gls{sssr} streamer}~\cite{Scheffler2023SparseSS} provides two indirection-capable \glspl{sr} (\texttt{ft0} and \texttt{ft1}) and another affine \gls{sr} (\texttt{ft2}).
The \emph{\gls{frep} hardware loop}~\cite{Zaruba2020SnitchAT} provides a short repetition buffer for offloaded \gls{fpu} instructions, allowing the integer processor and \gls{fpu} to concurrently execute instructions in a \emph{pseudo-dual-issue} fashion. 

The Snitch cluster also provides \SI{128}{\kibi\byte} of \gls{tcdm} across 32 banks; this memory is explicitly managed and enables high-bandwidth, low-latency accesses at \SI{64}{b} granularity from all eight cores simultaneously. 
A \SI{512}{\bit} programmable \gls{dma} engine~\cite{Benz2023AHE} enables high-bandwidth bulk data transfers between \gls{tcdm} and main memory.

\Cref{tbl:impl_saris_codes} lists the stencil codes we implement on the Snitch cluster and their performance-relevant characteristics, sorted by the number of \glspl{flop} per grid point.
All codes use double-precision data.
For each, we implement one time iteration on a $64^2$ (2D) or $16^3$ (3D) grid tile including halos.
We double-buffer the grid tiles in \gls{tcdm} and program the \gls{dma} engine to concurrently transfer them from and to main memory in 2D or 3D transfers as is required for our scaleout in \Cref{subsec:eval_scaleout}. Like our target platform, our codes are available open-source.

Each code is implemented in two parallelized variants, which are both compiled using a Snitch-optimized LLVM 15 toolchain.
The \emph{optimized baseline} (\textsc{base}) variants target the RV32G architecture without extensions; they are compiled using the \texttt{-Ofast} optimization level and a custom reassociation pass to maximize \gls{fpu} utilization.
The \emph{SARIS-accelerated} (\textsc{saris}) variants target the RV32G architecture with the \gls{sssr} and \gls{frep} extensions; they also use \texttt{-Ofast}, but have manually scheduled point loops using SARIS and use FREP where possible.
Both \textsc{base} and \textsc{saris} variants parallelize their point loops among the eight cluster cores using four-fold $x$-axis and two-fold $y$-axis iteration interleaving, and further unroll their point loops up to four-fold \emph{iff} beneficial to performance.



\section{Evaluation}
\label{sec:eval}

We first discuss the performance (\Cref{subsec:eval_perf}) and energy efficiency benefits (\Cref{subsec:eval_enpow}) of SARIS on a single eight-core cluster.
We then estimate the performance benefits of SARIS in a 256-core manycore scaleout (\Cref{subsec:eval_scaleout}).

\subsection{Performance}
\label{subsec:eval_perf}

\begin{figure}[t]
  \centering
  \subfloat[Execution speedup of \textsc{saris} over \textsc{base} code variants.]{
    \label{fig:res_perf_su}
    \includegraphics[width=0.975\linewidth]{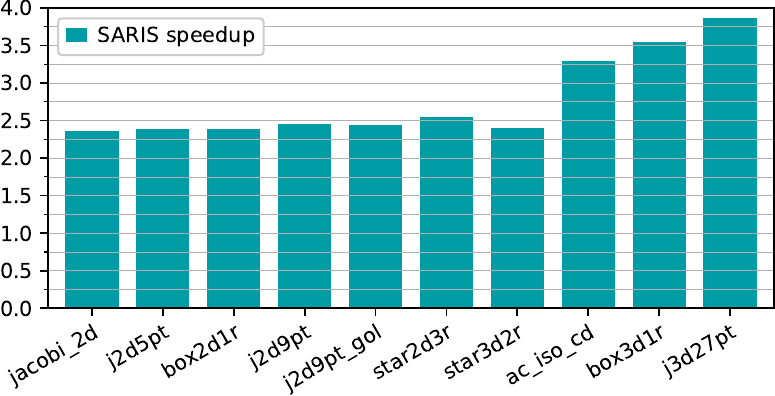}
  }%
  \vspace{1em}
  \captionsetup[subfigure]{justification=centering}
  \subfloat[\gls{fpu} utilization and per-core \glsunset{ipc}\gls{ipc} for both  variants.]{
    \label{fig:res_perf_fpipc}
    \includegraphics[width=0.975\linewidth]{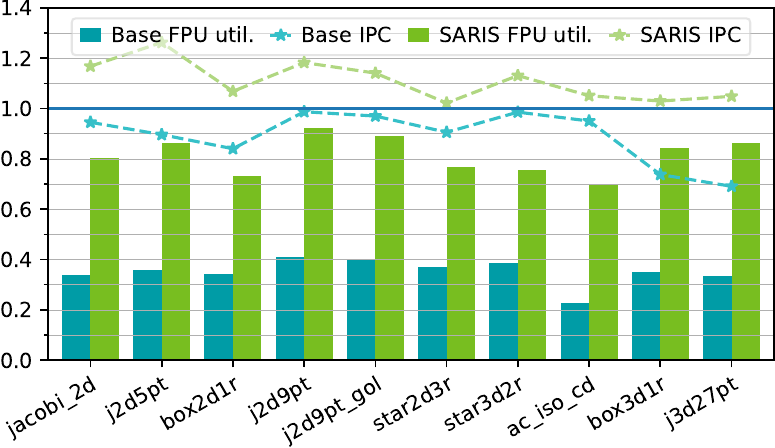}
  }
  \caption{Performance comparison of \textsc{base} and \textsc{saris} stencil code variants, with codes sorted by \glspl{flop} per grid point.}
  \label{fig:res_perf}
\end{figure}

We execute our stencil codes in cycle-accurate simulations of the Snitch cluster's \gls{rtl} description. We then extract exact runtimes and utilization metrics from simulation traces.

\Cref{fig:res_perf_su} shows the speedups of \textsc{saris} over \textsc{base} code variants.
SARIS achieves a significant geomean speedup of \impval{2.72\x}, with a clear increasing trend as \glspl{flop} per grid point increase;
the codes with the fewest (\texttt{jacobi\_2d}) and most (\texttt{j3d27pt}) \glspl{flop} per grid point also exhibit lowest and highest speedups of \impval{2.36\x}~and \impval{3.87\x}, respectively. 
In between, speedups remain approximately stable up to \texttt{star3d2r} (geomean \impval{2.42\x}), then increase further.

To explain this trend, we observe that codes past this point involve many \emph{grid loads} and \emph{coefficients} (see \Cref{tbl:impl_saris_codes}), increasing the register pressure in \textsc{base} variants. 
This reduces the benefits of unrolling, which may exhaust architectural registers and require inefficient stack accesses. 
However, reducing unrolling increases dependency stalls, also negatively impacting performance. 
SARIS avoids this register bottleneck by streaming grid points and register-exhausting coefficients directly from \gls{tcdm}, maintaining the full benefits of unrolling and further increasing speedups.

\Cref{fig:res_perf_fpipc} depicts the \gls{fpu} utilization and per-core \glsunset{ipc}\gls{ipc} for both variants. 
As expected from our example in \Cref{subsec:impl_saris}, SARIS significantly improves the geomean \gls{fpu} utilization from \impval{\SI{35}{\percent}} to a near-ideal \impval{\SI{81}{\percent}}. 
It also enables the effective use of Snitch's pseudo-dual-issue capabilities, increasing geomean \gls{ipc} from \impval{0.89} to \impval{1.11}.
SARIS performance remains consistently high, with \gls{fpu} utilization and \gls{ipc} never dropping below \impval{\SI{70}{\percent}} and \impval{1.0}, respectively.
Variations in \gls{fpu} utilization and \gls{ipc} can be attributed to multiple effects. 
For non-register-bound codes (up to \texttt{star3d2r}), \gls{ipc} variations in both variants resemble each other as both are impacted by \gls{tcdm} \emph{access contention}; codes with many \emph{grid loads} or large \emph{stencil overlaps} between cores exhibit more stalls due to bank conflicts.
For register-bound codes (past \texttt{star3d2r}), the \textsc{base} \gls{ipc} drops down to a minimum of \impval{0.69} due to dependency stalls, while \textsc{saris} variants avoid the register bottleneck through streaming as previously explained.
Finally, as the number of \emph{grid loads} and the \emph{stencil radius} increase, so does the setup overhead of SARIS: more indices must be stored for fewer point iterations doing useful compute. This is why \texttt{ac\_iso\_cd}, having the largest radius and many point loads, exhibits the minimum \textsc{saris} \gls{fpu} utilization of \impval{\SI{70}{\percent}}.

Overall, the inefficiencies preventing full \gls{fpu} utilization in \textsc{saris} codes are index initialization overheads, \gls{tcdm} access contention, instruction cache misses, and core runtime imbalances.

\subsection{Energy and Power}
\label{subsec:eval_enpow}

\begin{figure}[t]
  \centering
  \includegraphics[width=0.975\linewidth]{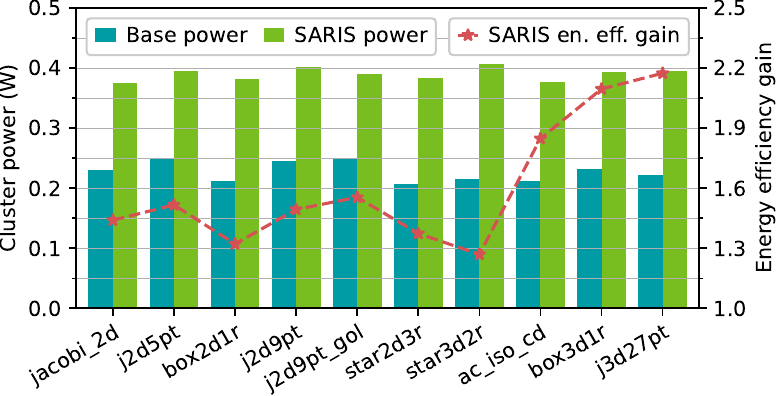}%
  \caption{Cluster power consumption for \textsc{base} and \textsc{saris} variants and \textsc{saris} energy efficiency improvement over \textsc{base}.}
  \label{fig:res_enpow}
\end{figure}

We implement the Snitch cluster in  GlobalFoundries’ 12LP+ FinFET technology using \emph{Fusion Compiler} and execute all considered stencil codes in post-layout gate-level simulation at the cluster's target clock speed of \SI{1}{\giga\hertz}.
We record the resulting switching activity and use it for cluster power estimation in \emph{PrimeTime}, assuming typical operating conditions of \SI{25}{\celsius} and \SI{0.8}{\volt} core supply voltage.

\Cref{fig:res_enpow} shows the resulting power consumption for each code, as well as the energy efficiency benefits of SARIS.
The power consumption for both variants is fairly stable across codes with geomeans of \impval{\SI{227}{\milli\watt}} and \impval{\SI{390}{\milli\watt}} for \textsc{base} and \textsc{saris}, respectively.
As would be expected, the slight variations in power consumption among either variant resemble those in \gls{fpu} utilization.
While the increased \gls{fpu} utilization of \textsc{saris} variants results in \impval{1.72\x}~higher mean power consumption, the significant speedups of SARIS result in notable energy efficiency gains for all codes, ranging from \impval{1.27\x}~to \impval{2.17\x}~with a geomean of \impval{1.58\x}.
Accordingly, we observe the same increase in energy efficiency gains for register-bound codes as for speedups.

\subsection{Manycore Scaleout}
\label{subsec:eval_scaleout}

\begin{figure}[t]
  \centering
  \includegraphics[width=0.975\linewidth]{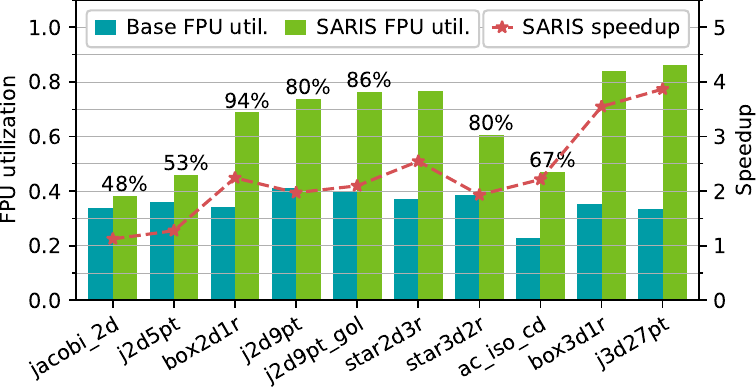}
  \caption{\gls{fpu} utilization for both code variants and speedup of \textsc{saris} over \textsc{base} in the scaled-out system. The compute-to-memory time ratio is shown for memory-bound stencils.}
  \label{fig:res_scaleout}
\end{figure}

We estimate the performance benefits of SARIS at scale on a Snitch-based manycore system.
We assume a simplified version of the Manticore~\cite{Zaruba2021Manticore} architecture we call \emph{Manticore-256s}, coupling one compute chiplet to one HBM2E memory stack with eight \impval{\SI{3.2}{\giga\bit\per\second\per pin}} devices.
The compute chiplet features one manager core and eight groups of four Snitch clusters each, totalling 256 cluster cores. Each group shares the bandwidth provided by one HBM2E device.

As in \cite{Matsumura2020AN5DAS}, we use grid sizes of \impval{$16384^2$} for 2D and \impval{$512^3$} for 3D codes.
We assume the mean \gls{dma} bandwidth utilization measured in our single-cluster experiments and that available group bandwidth is shared equally among clusters.
We model the imbalance in tile computation times due to scheduling variations and contention among clusters by assuming the same distribution for runtime imbalance among clusters as we observe among cores in a cluster.

\Cref{fig:res_scaleout} shows the estimated \gls{fpu} utilizations for the \textsc{base} and \textsc{saris} variants, as well as the estimated SARIS speedups. 
For memory-bound codes, we also indicate the \gls{cmtr}.
Despite entering the memory-bound regime for \impval{seven} out of \impval{ten} codes, SARIS improves the geomean FPU utilization from \impval{\SI{35}{\percent}} to \impval{\SI{64}{\percent}} and achieves a \impval{2.14\x} geomean speedup over \textsc{base} variants, reaching a peak performance of \impval{\SI{406}{\giga\flop\per\second}}.

For the \textsc{base} variants, \gls{fpu} utilizations closely mirror the single-cluster results from \Cref{fig:res_perf_fpipc}, as all codes remain compute-bound despite bandwidth sharing among clusters.
For the \textsc{saris} variants, the memory-bound codes see notable degradations in their performance.
As would be expected, codes with few \glspl{flop} per grid point exhibit a low operational intensity and thus a low \gls{cmtr}, making them memory bound. 
As the \glspl{flop} per grid point rises, so does the \gls{cmtr} until workloads are no longer memory-bound.
For the least operationally intensive 3D codes (\texttt{star3d2r} and \texttt{ac\_iso\_cd}), we see a regression to memory-boundedness; this is because 3D halos more strongly reduce on the ratio of input to output points in a tile, further impacting operational intensity.
Moreover, \texttt{ac\_iso\_cd} accesses additional I/O arrays for a prior time step and a time-dependent impulse, further increasing its memory-boundedness.

Nonetheless, while SARIS yields its greatest benefits in compute-bound scenarios, it notably enhances performance even for memory-bound stencils, which, in our experiments, still show speedups as high as \impval{2.25\x} over \textsc{base} variants and a geomean speedup of \impval{1.78\x}.


\section{Related work}
\label{sec:relwork}

While some \gls{sr} proposals like \emph{stream semantic registers}~\cite{Schuiki2021StreamSR} and the \emph{unlimited vector extension}~\cite{Domingos2021UnlimitedVE} accelerate selected stencil benchmarks using affine streams, 
to the best of our knowledge, SARIS is the first generic stencil acceleration method leveraging indirect \glspl{sr} to enable near-ideal FPU utilizations.
Beyond \glspl{sr}, numerous existing works accelerate stencil codes with dedicated hardware or through optimized CPU, GPU, and \gls{wse} software:

\topar{Hardware Accelerators} dedicated stencil accelerators often target \glsunset{fpga}\glspl{fpga} or operate near memory.
\emph{Casper}~\cite{Denzler2021CasperAS} accelerates stencils near a CPU's \gls{llc} by leveraging its high bandwidth and data locality, incurring minimal area.
\emph{Nero}~\cite{Singh2020NEROAN} is a CAPI2-interfaced \gls{fpga}+HBM solution using high-level synthesis to accelerate two compound kernels fundamental to weather prediction.
\emph{SODA}~\cite{Chi2018SODASW} automatically generates optimized dataflow architectures for stencil algorithms on \glspl{fpga} with provably optimal data reuse.

\topar{CPUs} Many CPU software approaches improve stencil performance through \glspl{dlt} and vectorization.
Zhang et al.~\cite{Zhang2020DataLT} vectorize stencil codes for ARM's NEON extension by packing grid loads for multiple output points contiguously in memory, reaching up to \impval{\SI{29}{\percent}} of peak compute on one core of an FT-2000+ processor.
Yount~\cite{Yount2015VectorFI} proposes \emph{vector folding}, a \gls{dlt} forming $n$D \emph{folds} in memory to vectorize stencils along multiple axes, using up to \impval{\SI{30}{\percent}} of a Xeon Phi 7120A's peak compute.
\emph{Bricks}~\cite{Zhao2018DeliveringPS} reorganizes large grids into fine-grain blocks contiguous in memory to exploit the full $n$D data locality of stencils and reduce \glsunset{tlb}\gls{tlb} pressure, achieving up to \impval{\SI{45}{\percent}} of peak compute on Xeon Gold 6130.

\topar{GPUs} Numerous highly optimized code generators have been proposed for stencils on GPUs.
\emph{ARTEMIS}~\cite{Rawat2019OnOptimizingCS} leverages many existing and novel tiling, decomposition, fusion, and folding optimizations with parameters guided by profiling; it enables up to \impval{\SI{36}{\percent}} of peak performance on a P100 GPU.
\emph{DRStencil}~\cite{You2021DRStencilED} improves operational intensity and data reuse by fusing time steps and partitioning the resulting stencils with autotuned parameters, achieving up to \impval{\SI{48}{\percent}} of peak compute on P100.
\emph{AN5D}~\cite{Matsumura2020AN5DAS} uses a performance model to generate highly optimized CUDA code with automatic spatial and temporal blocking and \gls{sm} double-buffering from a C source, improving performance scaling and further increasing peak compute utilization to \impval{\SI{69}{\percent}} on V100 SXM2.
\emph{EBISU}~\cite{Zhang2023RevisitingTB} leverages the much larger \gls{sm} in newer GPUs by trading deep temporal blocking for lower device occupancy, achieving notable speedups over AN5D despite using only \impval{\SI{49}{\percent}} of peak compute on A100.

\topar{Wafer-Scale Engines} the spatial architecture and high on-chip bandwidth of \glspl{wse} makes them attractive for stencil computation scaleouts on large grids.  
Rocki et al.~\cite{Rocki2020FastSC} solve a large linear system arising from a 7-point stencil with mixed precision on Cerebras WSE-1, utilizing \impval{\SI{28}{\percent}} of the system's peak performance.
Jacquelin et al.~\cite{Jacquelin2022ScalableDH} scale out a 25-point acoustic isotropic constant-density (\texttt{ac\_iso\_cd} in our evaluation) seismic simulation kernel on the even larger WSE-2, mapping grid planes to tiles and streaming along the $z$ axis; they use \impval{\SI{28}{\percent}} of their engine's peak compute.

\Cref{tbl:relwork_sw} summarizes the discussed software works and compares the highest fraction of peak compute reported by each. 
We see that SARIS on our Manticore-256s scaleout reaches the highest peak performance utilization, \impval{\SI{15}{\percent}} higher than the leading GPU code generator \emph{AN5D}.
Methodologically, most of the discussed software innovations are orthogonal to SARIS; while some \gls{dlt} approaches may be obviated by the fine-grain gathering abilities of indirect \glspl{sr}, further optimizations improving data reuse, arithmetic intensity, locality and tiling could all benefit our Manticore-256s scaleout with SARIS, potentially enabling even higher performance.

\begin{table}
\centering
\small
\def\arraystretch{0.91}%
\begin{tabular}{cp{8.5em}p{8.5em}cc}
\toprule
& Work &%
Platform &%
Prec. &%
\% Pk. \\
\midrule
\multirow{3}{*}{\adjustbox{angle=90}{CPU}}%
& Zhang et al.~\cite{Zhang2020DataLT}            & FT-2000+ (1 core) & FP64    & \SI{29}{\percent} \\
& Yount~\cite{Yount2015VectorFI}                 & Xeon Phi 7120A    & FP32    & \SI{30}{\percent} \\
& \emph{Bricks}~\cite{Zhao2018DeliveringPS}      & Xeon Gold 6130    & FP32    & \SI{45}{\percent} \\
\midrule
\multirow{4}{*}{\adjustbox{angle=90}{GPU}}%
& \emph{ARTEMIS}~\cite{Rawat2019OnOptimizingCS}  & Tesla P100        & FP64    & \SI{36}{\percent} \\
& \emph{DRStencil}~\cite{You2021DRStencilED}     & Tesla P100        & FP64    & \SI{48}{\percent} \\
& \emph{AN5D}~\cite{Matsumura2020AN5DAS}         & Tesla V100 SXM2   & FP32    & \SI{69}{\percent} \\
& \emph{EBISU}~\cite{Zhang2023RevisitingTB}      & A100              & FP64    & \SI{49}{\percent} \\
\midrule
\multirow{2}{*}{\adjustbox{angle=90}{WSE}}%
& Rocki et al.~\cite{Rocki2020FastSC}            & Cerebras WSE-1    & FP16-32 & \SI{28}{\percent} \\
& Jaquelin et al.~\cite{Jacquelin2022ScalableDH} & Cerebras WSE-2    & FP32    & \SI{28}{\percent} \\
\midrule
& \textbf{SARIS (Ours)}                          & Manticore-256s    & FP64    & \impval{\textbf{\SI{79}{\percent}}} \\
\bottomrule
\end{tabular}
\vspace{1em}
\caption{Overview of discussed stencil software approaches. \emph{\%~Pk.} is the highest fraction of peak compute achieved.}
\vspace{-1em}
\label{tbl:relwork_sw}
\end{table}


\section{Conclusion}
\label{sec:conclusion}

We present SARIS, a flexible and generic methodology to accelerate stencil 
codes with indirect \glspl{sr}. 
SARIS stores the offsets of grid point loads in static index arrays reused in each point iteration, using multiple indirect \glspl{sr} for concurrent operand streaming and additional affine \glspl{sr} to load coefficients or store results.
It works with any stencil shape, any number of I/O arrays, and is amenable to parallelization. 
It can be combined with code optimizations like loop unrolling, arithmetic reassociation, and the use of hardware loops.
We evaluate SARIS on the eight-core RISC-V Snitch cluster with \glspl{sssr} by implementing parallel optimized RV32G baseline and SARIS-accelerated variants of common stencil codes, which we make available open-source.
SARIS achieves significant speedups of \impval{2.72x}, near-ideal \gls{fpu} utilizations of \impval{81\%}, and notable energy efficiency improvements of \impval{1.58x} on average. 
Scaling our codes to the Manticore-256s manycore, SARIS enables an FPU utilization of \impval{\SI{64}{\percent}} and a \impval{2.14\x} speedup on average despite the bandwidth limitations of a complex memory system, and reaches up to \impval{\SI{15}{\percent}} higher fractions of peak compute than a leading GPU code generator.  

\begin{acks}
This work has been supported in part by funding from the European High-Performance Computing Joint Undertaking (JU) under Grant Agreement No 101034126 (The European Pilot) and Specific Grant Agreement No 101036168 (EPI SGA2).

\end{acks}


\bibliographystyle{ACM-Reference-Format}
\bibliography{main.bib}


\end{document}